# Unusual Field Dependence of Anomalous Hall Effect in Ta/TbFeCo

M.D. Davydova[1], Jong-Ching Wu,[2] Sheng-Zhe Ciou,[2] Yi-Ru Chiou,[2] P.N. Skirdkov[1,3], K.A. Zvezdin[1,3], A.V. Kimel[4], Lin-Xiu Ye,[5] Te-Ho Wu,[5] Ramesh Chandra Bhatt[5], and A.K. Zvezdin[3]

[1] Moscow Institute of Physics and Technology, Dolgoprudny 141701, Russia

[2] Department of Physics, National Changhua University of Education, Changhua 500, Taiwan

[3] Prokhorov General Physics Institute of the Russian Academy of Sciences, Moscow 119991, Russia

[4] Radboud University, Institute for Molecules and Materials, Nijmegen 6525 AJ, The Netherlands

[5] Graduate School of Materials Science, National Yunlin University of Science and Technology, Douliu, Yunlin 640 Taiwan

## Abstract

*Experimental studies of anomalous Hall effect are performed for thin filmed Ta/TbFeCo in a wide range of temperatures and magnetic fields up to 3 T. While far from the compensation temperature ($T_M$=277 K) the field dependence has a conventional shape of a single hysteresis loop, just below the compensation point the dependence is anomalous having the shape of a triple hysteresis. To understand this behavior, we experimentally reveal the magnetic phase diagram and theoretically analyze it in terms of spin-reorientation phase transitions. We show that one should expect anomalous hysteresis loops below the compensation point if in the vicinity of it the magnetic anisotropy is dominated by FeCo sublattice due to interaction with Ta.*

## Introduction

Rare-earth - transition metal amorphous alloys and intermetallics form a wide class of ferrimagnetic materials that is particularly interesting in spintronics[1,2], optospintronics[3], ultrafast magnetism[4,5], and magnonics[6]. One of their biggest advantages is the sensitivity of their magnetic and transport properties to a subtle change of the composition, temperature or application of magnetic field. GdFeCo and TbFeCo amorphous alloys are particular examples of such materials, and they have become model systems in ultrafast magnetism[7–11]. In the alloys 4*f* rare-earth (Gd, Tb) is coupled antiferromagnetically to the 3*d* transition metal (FeCo) sublattice. Both sublattices have the same Curie temperature but their magnetizations have different temperature dependencies. Therefore, by changing the concentrations of 4*f* and 3*d* elements in the compound the compensation temperature, at which the magnetizations of the sublattices are mutually equal, and the net magnetization is zero, can be tuned in a wide range of temperatures, including room temperature[12]. At temperatures lower than the compensation temperature, the magnetization of the rare-earth (Tb) sublattice $M_f$ is larger than that of the transition metal (FeCo) $M_d$, while for the temperatures above compensation point $M_f < M_d$. Ultrafast magnetization reversal in RE-TM materials was demonstrated at record-breaking rates across compensation temperature[13,14]. Also in the vicinity of the compensation temperature a relatively low magnetic field can turn the ferrimagnet into a noncollinear phase, in which the magnetic sublattices are canted and the net magnetization emerges[15]. Noncollinear spin configurations in ferrimagnets give rise to anomalous Hall effect (AHE)[16,17] and, thus, can be sensed in transport.

Earlier experimental studies of Tb(Fe$_x$Co$_{1-x}$)[18,19], Gd(Fe$_x$Co$_{1-x}$)[15,20–23], HoCo[24], DyCo4[25] revealed a rather unusual behavior of the magnetization in these compounds. Unlike a single hysteresis loop normally expected in measurements of the magnetization as a function of applied magnetic field, a triple hysteresis loops were observed. Unlike our reports about triple hysteresis observed in pure TbFeCo just above the compensation temperature $T_M$[19], here the situation is inverted that the triple hysteresis is observed below $T_M$. More particularly, using Hall bar structures of MgO/Ta/TbFeCo we measured field dependencies of the anomalous Hall effect at various temperatures with a special attention to domain in the vicinity of the compensation temperature. We experimentally define the magnetic phase diagram and propose a model that can explain this inversion. We show theoretically that the effect can result from an enhanced magnetic anisotropy of the *d*-sublattice. Such an enhancement can originate from Ta layer known for a strong spin-orbit interaction. Thus, we demonstrate that by changing the structure of the layered system one can significantly influence the composition of the magnetic phase diagram. We suggest that our findings are important from fundamental point of view as well as for designing novel RE-TM – based memory and spintronic devices with desired magnetic properties.

## Experimental technique and measurement results

Multilayers of MgO(4 nm)/Ta(5 nm)/TbFeCo(10 nm)/SiO$_2$-coated Si (100) were prepared by a high vacuum magnetron sputtering method. The magnetic Tb$_{28.5}$(Fe$_{80}$Co$_{20}$)$_{71.5}$ (the composition was identified using Inductively Coupled Plasma- Mass Spectroscopy) layer was fabricated using co-sputtering of a FeCo target and a Tb target. Sputtering power on each target was controlled for fine tuning of Tb concentration. No thermal annealing was performed. The Hall bar was fabricated as follows: (1) a standard photolithography was employed to delineate the pattern of outer electrodes and probing pads, an ion beam sputtering system equipped with two ion guns was then adopted for depositing copper film after using the second ion gun to etch the 4 nm MgO film on the top of the Ta/TbFeCo bilayer; (2) an electron beam lithography was used for defining the Hall bar shape device, followed by ion beam etching technique to transfer the Hall bar-shaped device. The Hall bars microscope image is shown in Fig. 1(a). A DC Hall measurement was carried out in a Physical Property Measurement System that is equipped with a superconducting magnet capable of applying magnetic field perpendicular to the film plane.

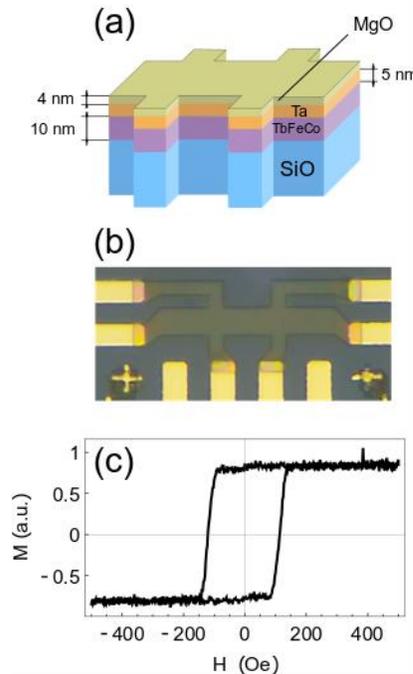

**Fig. 1.** (a) Scheme of the structure with layers, (b) optical microscope image of the Hall bar device. (c) Alternating gradient magnetometer measurement of magnetization loop of the film at room temperature.

We measure the DC AHE resistance at sensing current $I$ = 100 μA to obtain the hysteresis curves of $R_{AHE}$ depending on the strength of the applied magnetic field H for a wide range of temperatures. The results of these measurements are shown in Fig.2. We see that the direction of the hysteresis loop inverts above the compensation temperature (compare curves at 263 and 283 K). Immediately below the compensation temperature we observe anomalous behavior: triple hysteresis loops. They are shown for T = 273, 276.5 and 277 K on Fig.2(a) and in more details (T = 271, 272, 273, 274, and 276.5 K) on Fig. 2(b). At the magnetization compensation temperature $T_M$ the loops merge into one. We conclude that the magnetization (?) compensation temperature for the compound $T_M$ is between 276.5 and 277 K.

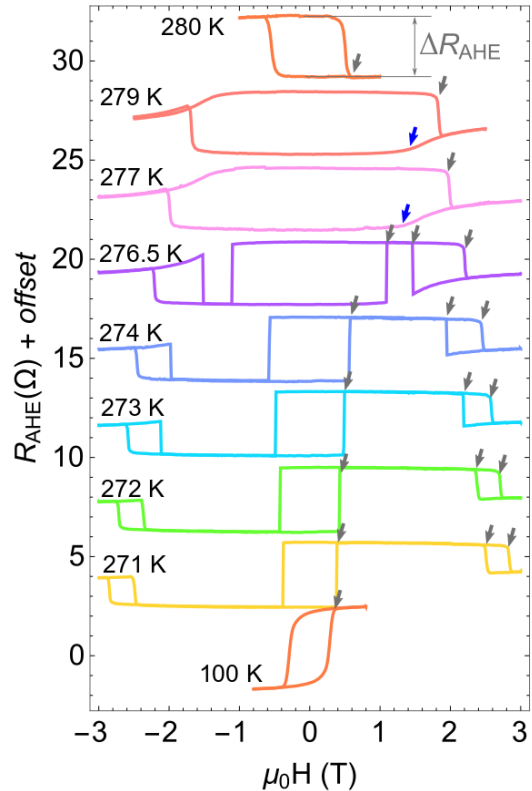

**Fig. 2.** Set of experimental AHE curves at different temperatures near the compensation point. The grey arrows indicate the edges of hysteresis at positive fields and the blue arrows point to the second-order phase transition points.

In Fig. 3(a) the hysteresis edges and spin-flop fields found using AHE measurements are shown for a wide range of temperatures. Grey diamonds correspond to the hysteresis edges that are illustrated by blue arrows in Fig. 2. Blue circles correspond to the spin-flop fields; examples of such transitions are located at blue arrows in Fig. 2. This transition is seen, for example, in Fig. 2 at T = 277 K as sharp bend in hysteresis curve at approximately 1.5 T. The growth of coercivity near 277 K confirms that it indeed is the compensation point.

In Fig. 3(b) the measured dependence of the height of the central hysteresis $\Delta R_{AHE}$ on temperature is shown. The sign flips at the compensation temperature. The AHE signal is sensitive mostly to one (FeCo). This is why the resistivity changes sign at the compensation temperature where the sublattice magnetization flips.

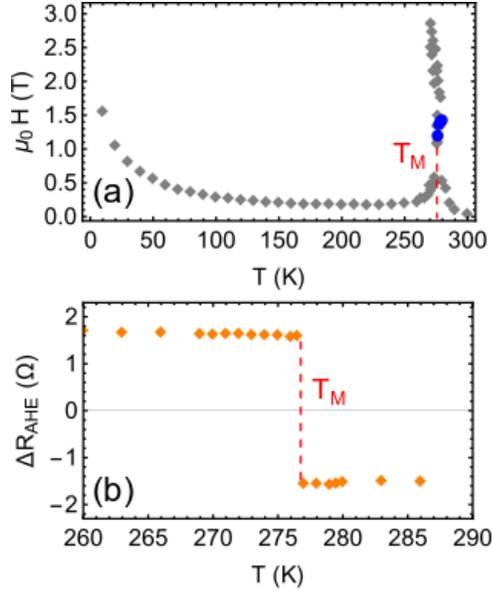

**Fig.3** The width i.e. coercive field (panel a) and the height $\Delta R_{AHE}$ (panel b) of the central hysteresis loop of the anomalous Hall effect as functions of temperature. The compensation temperature $T_M \approx 277$ K.

## Discussion

In order to examine the unusual behavior in more details we explore the peculiarities of the phase diagram in the vicinity of the compensation temperature. Figure 4 shows experimentally defined *H-T* magnetic phase diagram of the studied Ta/TbFeCo structure. Grey diamonds correspond to the edges of the hysteresis loops, and blue points correspond to the spin-flop transition points. The lines are derived from the theoretical explanation that is similar to refs. [19,26]. Similarly to our earlier reports, here we also interpret a hysteresis as a result of first-order phase transition. If a change is gradual, it must be assigned to second order phase transition.

We interpret the additional hysteresis loops as the consequence of the first-order phase transition at high fields, which is always surrounded by hysteresis. The spin-flop points correspond to a second-order phase transition. In the diagram, the grey dashed lines *BR*, *PR*, and *PA* correspond to the lines of the stability loss of different phases that surround the first-order transition lines between these phases, one at *H* = 0 and another one indicated by red line in Fig. 4. The blue solid lines (*RB'* and to the left from point *P*) are the second-order phase transitions; the point *P* is the tricritical point. Because of the additional first-order phase transition $T_MRP$ that bends towards lower temperatures with the increase of magnetic field, there exist anomalous hysteresis loops.

The temperature range of the first-order phase transition analogous to $T_MRP$ in crystalline media is usually very small, of the order of less than 1 K [27]. TbFeCo is an amorphous alloy, and in this case a sperimagnetic order[28] emerges near the phase transition points. In the sperimagnetic phase there exists a distribution of magnetic moments within a cone directed along the easy axis, which broadens the phase transitions, and, thus, might be the reason behind an increased temperature range of this first-order phase transition up to 10 K in the experiment. The sperimagnetism of rare-earth – transition metal compounds has been studied earlier[28–33].

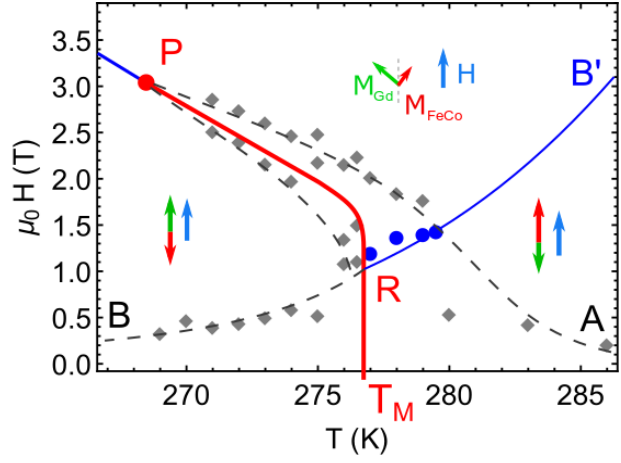

**Fig. 4** Experimentally measured features of the magnetic phase diagram near the compensation temperature. Gray and blue data points correspond to the edges of hysteresis and spin-flop fields, respectively. The orientation of magnetization vectors $M_{Tb}$ and $M_{FeCo}$ relatively to the external magnetic field $H$ in different phases is shown schematically by green and red arrows, correspondingly. The lines are the guides to the eyes that are drawn according to theoretical predictions (see below in the text).

To explain qualitatively the observed magnetic behavior, let us consider the following simple model. The free energy for a two-sublattice f-d (rare earth – transition metal) ferrimagnet with antiferromagnetic exchange has the form[34]:

$$F = -M_{FeCo} H \cos\theta_d - M_{Tb}\left(H^2 + (\lambda M_{FeCo})^2 - 2\lambda M_{FeCo} H \cos\theta_d\right)^{\frac{1}{2}} - K_u \cos\theta, \quad (1)$$

where the first term is the Zeeman energy of the FeCo sublattice, which is saturated by large d-d exchange, the second term is the energy of the rare-earth sublattice in the effective field that is a sum of external field $H$ and the intersublattice exchange field $H_{ex} = \lambda M_{FeCo}$. The latter is usually of the order of 50 T[34]. The last term in eq. (3) represents the anisotropy energy. The sublattices are coupled antiferromagnetically. $K_u$ is the uniaxial magnetic anisotropy constant that is a sum of several contributions: $K_u = -2\pi m^2 + K_{FeCo} + K_{Tb} + \frac{2K_s}{d}$. The first term is due to the demagnetizing field, $K_{FeCo,Gd}$ are the uniaxial anisotropy constants of FeCo- and Tb- sublattice, respectively, and $\frac{2K_s}{d}$ is the surface anisotropy introduced by Ta capping layer depending on the total thickness of the film $d$.

At the magnetization compensation point $T_M$ the sublattice magnetizations are equal $M_{Tb} = M_{FeCo}$. When $H \ll H_{ex}$, the only relevant term in the free energy is the anisotropy energy. The external magnetic field that is applied parallel along the easy magnetization axis leads only to renormalization of the anisotropy energy of the film. The energy can be written in the following form:

$$E_A \approx -\left(K_u - \frac{\chi_\perp H^2}{2}\right)\cos^2\theta \quad (2)$$

where $\theta$ is the polar angle defining the orientation of Neel vector $\mathbf{L} = \mathbf{M}_{FeCo} - \mathbf{M}_{Tb}$ (z-axis is parallel to the normal to the film), $\chi_\perp$ is the perpendicular magnetic susceptibility, $\chi_\perp \approx \frac{2M}{H_{ex}} \gg \chi_\parallel$, $M = M_{FeCo} \approx M_{Tb}$. At the value of the magnetic field $H = H^* = (H_A H_{ex})^{1/2}$, the effective anisotropy constant $-\left(K_u - \frac{\chi_\perp H^2}{2}\right)$ changes sign, which is the reason of a magnetic spin-flop transition. At the transition point, the angle $\theta$ changes in a jump-like fashion from $\pi$ to $\frac{\pi}{2}$. $H_A = 2K_u/M$ is the anisotropy field.

Let us now consider the temperatures below the compensation point $T < T_M$, where the first-order phase transition is observed in experiment. Earlier studies of rare-earth ferrimagnets have found that in the case of weak rare-earth magnetic anisotropy the transition shifts towards the lower temperatures with the increase of the magnetic field[35]. On the other hand, from more general considerations we have shown[19,26,36] that if the rare-earth anisotropy is prevailing an inclination of the first-order phase transition line towards higher temperatures would occur; therefore, we assume that the surface anisotropy introduced by Ta capping influences d-states that relatively more delocalized, and thus, for Ta-coated TbFeCo the f-sublattice anisotropy doesn't dominate at least in the vicinity of the compensation temperature of the film. The physical reason behind this is that Ta, electronic configuration of which is [Xe] $4f^{14}\,5d^3\,6s^2$, has large atomic number $Z = 73$ and spin-orbit coupling scales approximately as $Z^4$ [37]. Therefore, due to Ta coating the d-subsystem of the material has increased magnetic anisotropy.

At $T < T_M$, the sublattice magnetizations are not equal anymore and Zeeman energy emerges. Near the compensation point in the linear order of $H/H_{ex}$ it has the form[34]:

$$E_Z \approx -(M_{FeCo} - M_{Tb})\,H\cos\theta. \qquad (3)$$

This interaction stabilizes the collinear phase where $\theta_d = \theta = \pi$. Due to the competition between Zeeman energy (3) and anisotropy energy (2), a noncollinear phase emerges near the compensation point that is described by the angle $\theta_d = \theta_{NC}(H, T \approx T_M)$, which is close to $\pi/2$. In a certain neighborhood of the compensation temperature at $T < T_M$ and $H \geq H^*$ there exists an area where two phases, collinear $\theta_d = \theta_C = \pi$ and noncollinear one $\theta_d = \theta_{NC}(H, T)$, coexist. The area is defined by simultaneous satisfaction of the conditions of existence of the noncollinear phase the collinear phases. The situation when two such solutions coexist is illustrated in Fig. 5. The possible phases correspond to the local minima of the free energy (1). The phase transition between these two phases that occurs within this area is of the first order, which explains the observed triple hysteresis loops below the compensation point. Above the compensation temperature the condition for coexistence of the two phases is not satisfied anywhere and the second-order transition takes place.

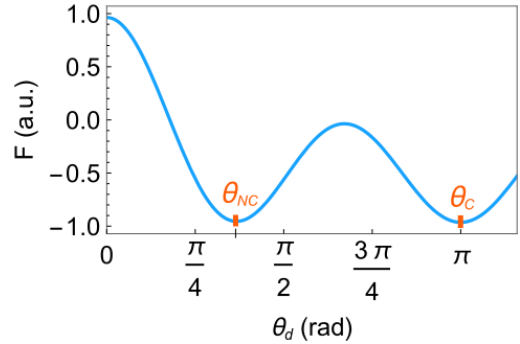

**Fig. 5** Dependence of free energy (1) on the angle $\theta_d$ below the compensation temperature at $M_{Tb}/M_{FeCo} = 1.002$, $H_A = 0.01\,H_{ex}$ and $H = 0.1 H_{ex}$, which corresponds to the vicinity of the first-order phase transition. The possible phases correspond to the minima of free energy.

## Conclusions

In this article, we reported unusual hysteresis loops of anomalous Hall effect in $Tb_{28.5}(Fe_{80}Co_{20})_{71.5}$ ferrimagnetic alloy. From the experimental data we deduce a magnetic phase diagram. We explain the loops in terms of spin-reorientation phase transitions. It is shown that the anomalous hysteresis loops will take place below the magnetization compensation if the magnetic anisotropy is dominated by the contribution from the transition metal sublattice. In contrast to pure TbFeCo the magnetic anisotropy experienced by the spins of the transition metal can be enhanced by the presence of the Ta capping known for strong spin-orbit interaction We believe that our findings offer a novel way for engineering the desired phase diagram of magnetic heterostructues.

## Acknowledgements

This research has been supported by RSF grant No. 17-12-01333.

# Supplementary Materials
to the article
# Unusual magnetic phase diagram in Ta/TbFeCo nanostructure revealed by AHE measurements

**S1. AHE measurements at wide range of temperatures**

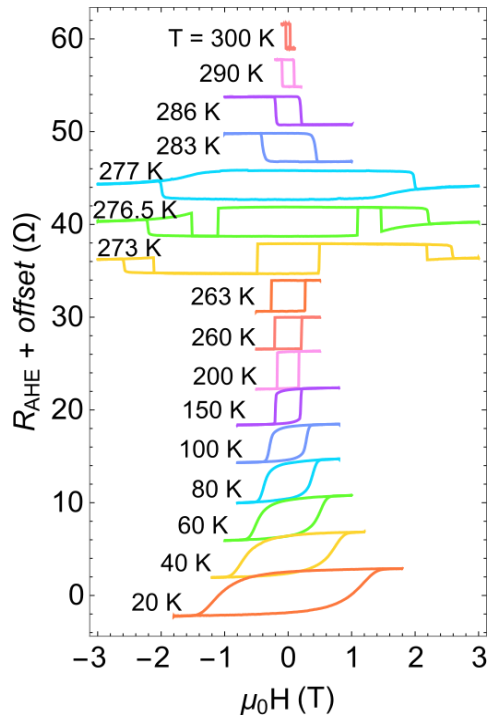

**Fig. S1.** Set of experimental AHE curves at different in a wide range of temperatures.